\documentclass[12pt]{article}

\author{Brett M. Savoie, Michael A. Webb, Thomas F. Miller III}
\title{\bf{Enhancing Cation Diffusion and Suppressing Anion Diffusion via Lewis-Acidic Polymer Electrolytes}}

\usepackage{mathtools}
\usepackage{graphicx}
\usepackage{chngpage}
\usepackage[version=3]{mhchem} % Formula subscripts using \ce{}
\usepackage{amsmath}
\usepackage{geometry}
\usepackage[superscript]{cite}         % Superscript citations
\usepackage{blindtext, fancyhdr}
\usepackage{upgreek}
\usepackage{xcolor}
\usepackage[normalem]{ulem}
\usepackage{multicol}
\usepackage{setspace}
\begin{document}
\maketitle
\renewcommand{\arraystretch}{1.25}
\rhead[Savoie, Webb, and Miller - {\thepage}]{Savoie, Webb, and Miller - {\thepage}}
\lhead[]{}
\newcommand{\red}{\textcolor{red}}

%%%%%%%%%%%%%%%%%%%
% Referenced Opening Paragraph  %
%%%%%%%%%%%%%%%%%%%
% ~150 words for Nature Materials (letters don't have an abstract, rather a referenced introductory paragraph that should blend seamlessly into the introduction)
\textbf{
Solid polymer electrolytes (SPE) have the potential to increase both the energy density and stability of lithium-based batteries, but low \ce{Li+} conductivity remains a barrier to technological viability. \cite{Tarascon:2001aa,Gray:1997} SPEs are designed to maximize \ce{Li+} diffusivity relative to the anion, while maintaining sufficient salt solubility. It is thus remarkable that polyethylene oxide (PEO), the most widely used SPE, exhibits \ce{Li+} diffusivity that is an order of magnitude smaller than that of typical counter-ions, such as TFSI${}^-$, at moderate salt concentrations.\cite{Edman:2002aa,Hayamizu:2002aa} Here, we show that Lewis-basic polymers like PEO intrinsically favor slow cation and rapid anion diffusion while this relationship can be reversed in Lewis-acidic polymers. Using molecular dynamics (MD) simulations, Lewis-acidic polyboranes are identified that achieve up to a ten-fold increase in \ce{Li+} diffusivity and a significant decrease in anion diffusivity, relative to PEO. The results for this new class of Lewis-acidic SPEs illustrate a general principle for increasing \ce{Li+} diffusivity and transference number with polymer chemistries that exhibit weaker cation and stronger anion coordination.
}
\newpage

%%%%%%%%%
% Introduction  %
%%%%%%%%%
PEO-based materials are among the most successful SPEs, with net conductivities on the order of $10^{-4}-10^{-3}~\mathrm{S\cdot cm^{-1}}$ at ambient temperature.\cite{Wright:2015aa,Zhang:2008aa} 
However, net conductivity actually overstates electrolyte performance, since only \ce{Li+} typically participates in the electrode chemistries in lithium-based batteries (Fig. 1A).
Measurements of the \ce{Li+} transference number, $T_\mathrm{Li}$ (the ratio of \ce{Li+} conductivity to the total conductivity), show that anions are responsible for most of the conductivity in PEO,\cite{Gorecki:1995aa,Johansson:1996aa,Leveque:1985aa} and ion diffusivity measurements reveal that anion diffusivity is an order of magnitude larger than \ce{Li+} diffusivity for common salts at typical concentrations.\cite{Edman:2002aa,Hayamizu:2002aa} 
Most strategies for increasing $T_\mathrm{Li}$ have focused on immobilizing the anion\cite{Hardy:1985aa,McBreen:2000aa,Bouchet:2013aa,Reddy:2014aa}, while the primary strategy for increasing overall diffusion rates has been to decrease the glass-transition temperature, $T_\mathrm{g}$, of the polymer.\cite{Wright:2015aa,Zhang:2008aa}
However, neither approach addresses the fundamental ion-polymer interactions that are responsible for asymmetric cation and anion conduction. 

The major finding of this work is that the low \ce{Li+} diffusivity and high anion diffusivity that characterize PEO-based SPEs can be reversed to favor \ce{Li+} conduction in Lewis-acidic polymers.
In conventional SPEs based on polyethers and other Lewis-basic units, salt solubility is driven by strong cation-polymer interactions.\cite{Fenton:2014aa,Shriver:1981aa,McBreen:2000aa}
However, this preferential coordination of cations leads to both the high relative diffusivity of weakly coordinated anions and the low diffusivity of \ce{Li+} (Fig. 1B). 
The trade-off between strong cation coordination and diffusivity suggests that the strategy of driving salt solubility with relatively stronger anion-polymer interactions and weaker cation-polymer interactions may enhance SPE performance. 
To investigate this trade-off, we present over 100 microseconds of MD simulations
to characterize the ion diffusivities, coordination structures, and solvation free energies in PEO and four Lewis-acidic polyboranes (Fig. 1C).
The employed force-fields are parameterized from \emph{ab initio} electronic structure calculations (methods) without experimental fitting; this approach provides a consistent level of theory for all studied polymers and enables the description of the  Lewis-acidic polymers for which experimental data does not exist (See Supporting Information for a discussion of force-field validation).  

Throughout this study, we primarily focus on the dilute-ion regime, with additional finite-concentration results presented in the supporting information.
%
%To exclusively characterize the effects of varying ion-polymer interactions, this work focuses on the dilute-ion concentration regime.
Consideration of the dilute-ion regime enables isolation of the ion-polymer interactions that are responsible for the solvation and transport of each ion,\cite{Bruce:1993aa,Webb:2015aa} while avoiding ion-pairing effects\cite{Borodin:2006aa,Watanabe:1987aa}
 and increases in polymer viscosity\cite{Shriver:1981aa,Gray:1997} that arise at higher ion concentrations.
For many polymer-salt combinations, the relative ion-diffusivity in the dilute-ion regime correlates strongly with results obtained at higher concentrations.\cite{Pesko:2016gf,Timachova:2015aa,Barteau:2013aa,Webb:2015aa} 
%However, we note that for other cases, such as LiCl in ether-based polymers and solvents, the dilute-ion regime is difficult to access experimentally due to extensive ion-pairing, even at low concentrations.\cite{Watanabe:1987aa,Muhuri:1997aa}
  
%%%%%%%%%
% PEO Results %
%%%%%%%%%
To illustrate the relative \ce{Li+} and anion diffusivities in a Lewis-basic environment, Fig. 1B presents MD simulations of ion diffusion in PEO in the dilute-ion regime. 
Results are presented for several anions, including \ce{Cl-}, triflate, and TFSI; these anions vary in both size and charge delocalization, with larger anions possessing increased charge delocalization and reduced Li-salt lattice energies.\cite{Ratner:1988aa,Borodin:2006aa}
Comparison of the cation and anion diffusivities in PEO reveals that the anion diffusivities dramatically exceed that of \ce{Li+} by eight- to thirty-fold, depending on the anion.  
For PEO:LiTFSI, the simulated dilute ion concentration diffusivities presented in Fig. 1B and Table 1 ($7.9\times10^{-8} \, \mathrm{cm}^{2} \cdot \mathrm{s}^{-1}$ for \ce{Li+} and $5.9\times10^{-7} \, \mathrm{cm}^{2} \cdot \mathrm{s}^{-1}$ for TFSI) show good agreement with NMR-based diffusivity measurements at similar temperatures and at dilute concentrations ($7\mathrm{-}9\times10^{-8} \, \mathrm{cm}^{2} \cdot \mathrm{s}^{-1}$ for \ce{Li+} and $4\mathrm{-}5\times10^{-7} \, \mathrm{cm}^{2} \cdot \mathrm{s}^{-1}$ for TFSI)\cite{Edman:2002aa}.
Direct comparison between theory and experiment for  PEO:LiCl and PEO:LiTriflate  in Fig. 1B is not possible, as ion-pairing occurs even at the lowest concentrations that have been experimentally studied.\cite{Watanabe:1987aa,Muhuri:1997aa,Boden:2002aa}
%

%while for LiCl in ether-based polymers and solvents extensive ion-pairing suppresses \ce{Cl-} diffusion even at dilute ion concentrations.\cite{Watanabe:1987aa,Muhuri:1997aa}}
%Fig. 1B demonstrates the dramatic difference in the cation and anion diffusivities in PEO, with anion diffusivities exceeding that of \ce{Li+} by approximately four- to ten-fold, depending on the anion.

Fig. 2 examines the molecular basis for the preferential anion diffusion in PEO. 
Fig. 2a (top row) presents representative coordination structures of all ions in PEO, and Fig. 2b (top panel) shows histograms of the number of polymer atoms that coordinate each ion. 
%In the PEO simulations, \ce{Li+} is helically coordinated by a single contiguous polymer segment with four oxygens in the equatorial positions and two more loosely associated oxygens in the axial positions, in good agreement with experimentally determined X-ray structures.\cite{Andreev:2005aa} 
Strong coordination of \ce{Li+} in PEO is reflected in the helical distortion of the polymer structure about the ion and the strongly peaked distribution of coordination structures (Fig. 2a). 
In contrast, each anion is weakly associated with a large number of methlyene units from several PEO segments, and a broad distribution of coordination structures is observed (Fig. 2b).  
These results demonstrate the intrinsic disadvantage that the coordination characteristics of PEO-based polymers represent to increasing \ce{Li+} diffusivity and $T_\mathrm{Li}$. 

%%%%%%%%%%%%%
% LA Solvation Results %
%%%%%%%%%%%%%
Additionally, Fig. 2 reveals that the asymmetric coordination of cations and anions in PEO is reversed in the Lewis-acidic polyboranes.
%Figs. 2a and 2b also present dilute-ion simulation results that indicate the asymmetric cation and anion coordination behavior observed in PEO is reversed in Lewis-acidic polyboranes.
Fig. 2a (left column) shows that the charge distributions of Lewis-acidic polymers is essentially inverted with respect to PEO; whereas PEO exhibits large negative charges that are localized on the oxygen and delocalized positive charges on the carbon backbone, the Lewis-acidic polymers exhibit large positive charges that are localized on the boron and delocalized negative charges on the remaining polymer atoms.
This inversion of the charge distribution between PEO and the Lewis-acidic polymers is reflected in the weaker coordination of \ce{Li+} and stronger coordination of the anions in the ion-coordination structures (Fig. 2a) and coordination-structure distributions (Fig. 2b).
Weak coordination of \ce{Li+} in the Lewis-acidic polymers is indicated by the larger number of polymer atoms coordinated with \ce{Li+} (Fig. 2a) and broader distribution of coordination structures (Fig. 2b), relative to PEO.
In contrast, the anion coordination structures in the Lewis-acidic polymers include fewer polymer atoms and narrower distributions.
Analysis of the ion-polymer radial-distribution functions (Fig. S1) also demonstrates the same trend.
These results indicate stronger anion coordination and weaker cation coordination in the Lewis-acidic polymers.  
%The combined observations from the dilute simulations indicate that the Lewis-acidic polymers exhibit increased anion coordination and decreased cation coordination relative to PEO. 

%%%%%%%%%%%%%%
% Diffusion Behavior - Li %
%%%%%%%%%%%%%%
Fig. 3 demonstrates that weaker coordination of \ce{Li+} in the Lewis-acidic polymers substantially improves the \ce{Li+} diffusivity, relative to PEO.
In two of the Lewis-acidic polymers, CBC and HBCC, \ce{Li+} diffusivity is increased four- to ten-fold, respectively. 
A third polymer, HBC, shows comparable \ce{Li+} diffusivity to PEO, while CBCC shows suppressed \ce{Li+} diffusivity. 
%To determine if the increased diffusivity is caused by more diffusive segmental motion of these polymers or if the solvation structure is more labile, the average duration of the major polymer-ion contact (Fig. 3b) and segmental mean-squared displacements (MSD) (Fig. 3c) were calculated for each system. 
To confirm that the increased \ce{Li+} diffusivity is primarily cause by increased lability of the coordination structures and not increased polymer segmental motion, polymer-ion contact autocorrelation functions $\langle h(0) h(t) \rangle$ (see Methods, Fig. 3b) and monomer-unit mean-squared displacement (MSD) (Fig. 3c) were calculated for each system. 
The data presented in Fig. 3b confirm that the shorter duration of the polymer-ion contacts correlates strongly with the observed \ce{Li+} diffusion. 
The monomer-unit MSDs shown in Fig. 3c are less correlated with the \ce{Li+} transport, although we note that local polymer fluctuations are important for facilitating transitions of \ce{Li+} among coordination structures.\cite{Borodin:2006aa,Webb:2015ab} 
%The results for segmental diffusion exhibit typical Rouse-like behavior for each polymer without any strong correlation with the \ce{Li+} transport. 
Notably, in both CBC and HBCC the \ce{Li+} MSD significantly exceeds the monomer-unit MSD, which indicates that the primary mechanism of \ce{Li+} transport is due to changes in coordination and not due to coupled diffusion with individual polymer segments.
%
%consistent with \ce{Li+} diffusion being decoupled from the diffusion of individual polymer segments on the simulated timescale and also characteristic of \ce{Li+} changing coordination. 

%%%%%%%%%%%%%%%%
% Diffusion Behavior - Anions %
%%%%%%%%%%%%%%%%
Similarly, Fig. 4 demonstrates that stronger coordination of \ce{Cl-} in the Lewis-acidic polymers substantially reduces the \ce{Cl-} diffusivity, relative to PEO.
Fig. 4a shows that the rapid \ce{Cl-} diffusion observed in PEO is suppressed in all of the Lewis-acidic polymers. 
Figs. 4b and 4c show the calculated polymer-\ce{Cl-} contact autocorrelation functions and polymer monomer unit MSDs, respectively. 
As in the case of \ce{Li+}, rapid diffusion coincides with short-lived polymer-\ce{Cl-} contacts, with the Lewis-acidic polymers all showing longer-lived polymer-\ce{Cl-} contacts relative to PEO. 
Fig. S2 contains the corresponding results for triflate and \ce{TFSI-}; in all cases longer-lived anion-polymer contacts correlate with decreased anion diffusivity, with most of the Lewis-acidic polymers showing longer-lived anion-polymer contacts in comparison with PEO.
The mechanism that emerges from these ion-diffusion simulations is that the Lewis-acidic polymers generally increase the strength of anion-polymer interactions and decrease the strength of \ce{Li+}-polymer interactions, leading to an increase in \ce{Li+} diffusion and relatively suppressed anion diffusion. 

%%%%%%%%%%%%%
%Transference Number %
%%%%%%%%%%%%%
Fig. 5a shows that in all cases, the Lewis-acidic polymers exhibit improved $T_{\mathrm{Li}}$ in comparison to PEO.
%Fig. 5a shows that the Lewis-acidic polymers, by facilitating weaker \ce{Li+} coordination and stronger anion coordination than PEO, exhibit systematically improved $T_{\mathrm{Li}}$.
The two Lewis-acidic polymers with the highest \ce{Li+} diffusivity, HBCC and CBC, also exhibit the largest improvements in $T_{\mathrm{Li}}$, reflecting that these polymers achieve both increased \ce{Li+} diffusivity and decreased anion diffusivity.
The magnitude of the improvement in $T_{\mathrm{Li}}$ also varies depending on the anion, with \ce{Cl-} showing the largest improvements in all Lewis-acidic polymers.
%The magnitude of the improvement in $T_{\mathrm{Li}}$ also varies depending on the anion, suggesting that the development of polymer-specific anions provides opportunity for further improvement.

%%%%%%%%%%%%%%%%%%
%Solvation Free Energies Number %
%%%%%%%%%%%%%%%%%%
The increased anion coordination in the Lewis-acidic polymers is also reflected in the relative contribution of the anion solvation to the total salt solvation.
Fig. 5b presents the ion-resolved solvation free energies associated with transferring each ion from vacuum into each polymer,
revealing that all combinations of Li-salt and Lewis-acidic polymer exhibit total solvation free energies that exceed the lattice energies of the salts. This suggests the capacity of the Lewis-acidic polymers to dissolve the studied salts, although the extent of ion-pairing cannot be established solely from dilute-ion simulations and is explored separately in Fig. S3. 
The capacity of the Lewis-acidic polymers to dissolve Li-salts is also supported by liquid-electrolyte studies; for instance, LiTriflate is soluble in borane-based Lewis-acidic solvents\cite{Zhang:2004aa} and both LiF and LiCl exhibit solubility increases of several orders of magnitude in liquid electrolytes when boranes are used as cosolvents.\cite{Lee:1998aa,Reddy:2014aa} 
Fig. 5c presents the fractional contribution of each anion to the total salt solvation free energy, demonstrating that the increases in $T_{\mathrm{Li}}$ are mirrored by increases in the relative contribution of the anion solvation free energy to the total salt solvation free energy. 
Taken together, the results in Fig. 5 suggest that high $T_{\mathrm{Li}}$ and high salt solubility can be simultaneously achieved by driving solubility with strong anion solvation while leaving \ce{Li+} only weakly complexed. 

Additional simulations performed at finite ion concentration indicate that, in some cases, the Lewis-acidic polymers exhibit comparable ion-pairing to PEO and increased conductivity (Fig. S3).   
Following previous work,\cite{Muller-Plathe:1994,Muller-Plathe:1994aa,Lin:2013aa,Lin:2013ab} 
the finite ion concentration simulations are performed with scaled partial charges on the ions to account for the effects of electronic polarizability, which reduces the viscosity and ion-pairing.  
This approach makes the finite ion concentration simulations more approximate than the dilute-ion simulations discussed earlier and prevents
the reliable comparison of finite ion concentration results among the different Li-salts, while still enabling  comparisons of ion-pairing and conductivity among polymers with a common salt. 
Fig. S3 shows that, in comparison to PEO, all of the Lewis-acidic polymers display increased $T_{\mathrm{Li}}$ with all salts and at nearly all concentrations. 
In particular, CBC:LiTFSI exhibits increased conductivity and reduced levels of ion-pairing compared to PEO:LiTFSI. %, as reflected by the degree of uncorrelated ion motion.

%%%%%%%%%%%%%%
% Concluding Statement %
%%%%%%%%%%%%%%
The current work represents a systematic computational study of ion transport in Lewis-acidic polymers and suggests a strategy for developing alternative polymer chemistries for lithium-based electrolytes. 
Dilute-ion simulations of PEO demonstrate that strong \ce{Li+} coordination and weak anion coordination manifest in dramatically suppressed \ce{Li+} transport relative to several common anions (Fig. 1b).
In contrast, in Lewis-acidic polymers, relatively weaker \ce{Li+} coordination and stronger anion coordination (Fig. 2) result in up to a ten-fold increase in \ce{Li+} diffusivity (Fig. 3) and twenty-fold increase in $T_{\mathrm{Li}}$ (Fig. 4a). 
The mechanism for the improved performance of the Lewis-acidic polymers is shorter-lived \ce{Li+}-polymer contacts (Fig. 3b) and longer-lived anion polymer contacts (Fig. 4b) that liberate \ce{Li+} diffusion relative to the anion in the Lewis-acidic polymers.
Comparisons of the ion-resolved solvation free energies in each polymer demonstrate that the Lewis-acidic polymers also exhibit lower \ce{Li+} sovation free energies (Fig. 4b) while generally increasing the relative contribution of the anion to the total salt solvation free energy (Fig. 4c).  
These results present a consistent picture of increasing SPE figures-of-merit via weakening \ce{Li+} coordination and strengthening anion coordination using Lewis-acidic polymer chemistries. 
Given the extremely slow pace of developing viable SPEs that are based on alternatives to polyether chemistry, the identification of a class of polymers that potentially overcomes the intrinsic limitations of PEO is encouraging.

PEO was the first polymer electrolyte discovered\cite{Fenton:2014aa} and is still the majority component of the highest-performing SPEs,\cite{Wright:2015aa,Bruce:2008} but new polymer chemistries are required to make further progress. 
Recent increases in \ce{Li+} diffusivity and conductivity have primarily been achieved through extrinsic innovations--the addition of nanoparticles and plasticizers,\cite{Croce:1998aa,Tarascon:2001aa} the use of low-lattice energy salts,\cite{Sylla:1992aa,Liang:2012aa} and the synthesis of amorphous, branched polymers\cite{Wright:2015aa,Zhang:2008aa}--while leaving the underlying ether-based polymer chemistry unchanged. 
Likewise, the main strategies for increasing $T_\mathrm{Li}$ include slowing anion diffusion via covalent immobilization\cite{Hardy:1985aa,Bouchet:2013aa}, Lewis-acidic additives\cite{McBreen:2000aa,Reddy:2014aa}, or dilute Lewis-acidic polymer moieties\cite{Lee:1999aa,Matsumi:2002p20952}, while leaving strong \ce{Li+} coordination in place. 

The current study demonstrates that the fundamental ion-polymer interactions responsible for both low $D_\mathrm{Li}$ and $T_\mathrm{Li}$ in PEO-based polymers can be favorably reversed in a Lewis-acidic environment.  
Although the detailed  study of the electrochemical decomposition of these materials is outside the scope of the current study,\cite{Barnes:2015ul}
prior uses of Lewis-acidic molecular additives\cite{McBreen:2000aa,Reddy:2014aa} suggest that the proposed polymer chemistries may be sufficiently stable for battery applications.
We also note that electrode interfacial impedance and dendrite growth pose significant materials challenges for the improvement of battery technologies,\cite{Tarascon:2001aa}
 although the  high \ce{Li+} diffusivity and $T_\mathrm{Li}$ for the Lewis-acidic polymers could potentially reduce these effects.\cite{Gray:1997}
The results presented here indicate that  removing  Lewis-basic units entirely or developing polymer chemistries with complementary weak Lewis-basic and strong Lewis-acidic moieties is a potentially more promising approach than optimizing $D_\mathrm{Li}$ and $T_{\mathrm{Li}}$ in the context of polyether chemistry. 
Likewise, developing anions with strong polymer-specific interactions is an opportunity for further improvement.  
The optimal balance will ultimately be determined by the solvating capacity and stability of the resulting polymer. 
\newline\newline
\section*{Acknowledgement} 
This research was supported by the National Science Foundation under DMREF Award Number NSF-CHE-1335486. 
M. A. W. also acknowledges support from the Resnick Sustainability Institute. 
This research used resources of the Oak Ridge Leadership Computing Facility at the Oak Ridge National Laboratory, which is supported by the Office of Science of the U.S. Department of Energy under Contract No. DE-AC05-00OR22725.
This research also used resources of the National Energy Research Scientific Computing Center, a DOE Office of Science User Facility supported by the Office of Science of the U.S. Department of Energy under Contract No. DE-AC02-05CH11231.

\section*{Methods}
\subsection*{Molecular Dynamics}
The molecular dynamics (MD) simulations employed a systematically parameterized version of the optimized potentials for liquid simulations (OPLS) force-field.\cite{OPLS1,OPLS2} 
 All intramolecular modes, atomic partial charges, and ion-polymer Lennard-Jones parameters were parameterized in this work to provide a consistent description of all polymers. 
Intramolecular modes, such as bonds, angles, and dihedrals, were parameterized by fitting OPLS potential energy terms to mode-scans performed at the B3LYP-D3/def2-TZVP level.
For each polymer, a model compound comprised of a tetramer of the polymer capped with methyl groups was employed for the intramolecular mode parametrization.
Approximate atomic partial charges were obtained from CHELPG calculations based on the B3LYP-D3/def2-TZVP electron densities at the optimized geometries.
After obtaining the intramolecular modes and approximate partial charges, condensed phase MD simulations were performed on the pure oligomers and also solutions with each salt to provide configurations to further refine the atomic partial charges and parametrize the dispersion interactions.
Final partial charges on each atom were determined by averaging the results of CHELPG calculations based on the B3LYP-D3/def2-TZVP electron densities for one hundred molecular configurations sampled from condensed phase MD trajectories.
Pairwise molecular configurations sampled from the condensed phase MD trajectories were used to parametrize the dispersion contributions to the polymer-polymer, ion-polymer, and ion-ion interactions. 
The dispersion interactions were parameterized by fitting Lennard-Jones potentials to the residual of the fixed electrostatic interactions and the counter-poise corrected B3LYP-D3/def2-TZVP interaction energies calculated for the pairwise configurations.
Full parametrization details, parameter tables, mode scans, and fit potentials are included in the Supporting Information. 

All MD simulations were performed within the LAMMPS software suite.\cite{Plimpton:1994aa} All trajectories employed periodic boundary conditions, particle-particle-particle-mesh (pppm) evaluations of long-range interactions beyond a 14 $\mathrm{\AA}$ cut-off, a Nos\'e-Hoover barostat with 1000 fs relaxation, and a Nos\'e-Hoover thermostat with 100 fs relaxation (NPT). 
Equations of motion were evolved using the velocity-Verlet integrator and a two femtosecond timestep for polymers without explicit hydrogen atoms and a one femtosecond timestep for polymers with explicit hydrogen atoms.
Intramolecular pairwise interactions for atom pairs connected by fewer than four bonds were excluded during the MD simulations to avoid double counting with dihedral energy terms.
Ion diffusion and solvation free energy trajectories were initialized from a common set of four independently equilibrated neat polymer trajectories. 
Each neat polymer trajectory included a single polymer chain with a mass of approximately 30 kDa that was initialized using a protocol to randomize chain orientation and avoid configurations with significant overlap between atoms. 
These configurations were initially equilibrated at a temperature of 10 K and a pressure of 50 atm for 50 ps, followed by a 10 ns annealing phase at a temperature of 500 K and a pressure of 1 atm.
After annealing, the configurations were simulated for 11 ns at a temperature of 400 K and a pressure of 1 atm to collect production data.
The first nanosecond of the production trajectories was used for equilibration and the remaining 10 ns were used to confirm the convergence of the density.
The final configurations from the neat polymer trajectories were used as input geometries for the ion diffusion and solvation free energy trajectories.

 \subsection*{Ion Diffusion Simulations}
For each combination of ion and polymer, sixteen independent ion-diffusion trajectories were performed.
These trajectories were initialized from the four pre-equilibrated neat polymer trajectories by randomly inserting a single ion into each configuration (four independent insertions for each pre-equilibrated configuration).
Each trajectory included a single ion to avoid correlated ion motions, and the excess charge was neutralized with a uniformly distributed background charge.
The initial geometry was relaxed by performing 1 ps of NVE dynamics with atom displacements limited to 0.01 $\mathrm{\AA}$ for each timestep, followed by 1 ns of NPT dynamics at a temperature of 400 K and pressure of 1 atm.
After relaxation, an additional 300 ns of NPT dynamics were performed to collect production data. 
A total of 4.8 $\mathrm{\mu}$s of ion diffusion dynamics were collected for each ion in each polymer (16 independent trajectories, each 300 ns long).  

Diffusivities can be calculated from the long-timescale trajectories using the Einstein equation
%\begin{equation}
%	\mathrm{D} = \frac{1}{6t} \left\langle \left| \mathbf{r}_{i}(t)-\mathbf{r}_{i}(0) \right|^{2}\right\rangle
%\end{equation}
\begin{equation}
	\label{diff_eq}
	D_i = \lim_{t\to\infty}\frac{\mathrm{d}\left\langle \left| \mathbf{r}_{i}(t)-\mathbf{r}_{i}(0) \right|^{2}\right\rangle}{6\mathrm{d}t} \, ,
\end{equation}
where $D_i$ is the diffusion coefficient for ion, $i$, and the term in brackets is the MSD evaluated at time $t$. 
Because many of the systems studied still exhibit sub-diffusive behavior even at long times, apparent diffusivities are reported by approximating the derivative in Eq.~\eqref{diff_eq} by finite difference using the MSD at $t=150$ and $t=0$~ns.\cite{Borodin:2006aa,Webb:2015aa} 
The \ce{Li+} transference number, $T_\mathrm{Li}$, was calculated using $T_\mathrm{Li}=D_\mathrm{Li}/(D_\mathrm{Li}+D_\mathrm{anion})$. 

Contact durations were calculated by defining a characteristic function, $h_{ij}(t)$, that reports on contacts between pairs of atoms
\begin{equation}
   h_{ij}(r) = \left\{
     \begin{array}{lr}
       1 & ,\mathrm{for}\,\, r_{ij} \leq r_\mathrm{T}\\
       0 & ,\mathrm{for}\,\, r_{ij} > r_\mathrm{T}
     \end{array}
   \right.  \, ,
\end{equation}
where $i$ and $j$ denote atomic indices, and $r_\mathrm{T}$ is a pair-specific threshold distance, which is chosen based on the size of the first coordination shell for the corresponding atom types.  
Specifically, $r_\mathrm{T}=r_\mathrm{max}+2\sigma$, where $r_\mathrm{max}$ is the radial separation at the first maximum of the corresponding  ion-polymer radial-distribution function (see Supporting Information), and $\sigma$ is the standard deviation obtained by fitting the full width at half maximum of the peak to a Gaussian function. 
The time autocorrelation function, $\left\langle h_{ij}(0)h_{ij}(t)\right\rangle$, was calculated for each contact type by averaging the trajectories with respect to time. Standard errors were calculated by separately averaging the results from the sixteen independent trajectories.
 
 \subsection*{Solvation Free-Energy Calculations}
Thermodynamic integration was used to calculate the ion-specific solvation free energies in each polymer.\cite{Frenkel:2001aa}
For each combination of ion and polymer, sixteen independent ion-insertion trajectories were used.
%These trajectories were initialized from the four pre-equilibrated neat polymer trajectories by randomly inserting a single ion into each configuration; the excess charge was neutralized with a uniformly distributed background charge.
%These trajectories were initialized by randomly inserting ions within the four pre-equilibrated neat polymer unit cells. 
Scaled potentials were used to gradually introduce the ion-polymer potential energy terms, and convergence was facilitated by introducing the polymer-ion Lennard-Jones interactions before introducing the polymer-ion electrostatic interactions. A scaled potential was used to first introduce the polymer-ion Lennard-Jones interactions
\begin{equation}
%  U_\mathrm{\lambda_1,LJ}(\lambda_1) =  U_\mathrm{P} + \lambda_1\left( U_\mathrm{P+X_{LJ}} - U_\mathrm{P} \right)
  U_\mathrm{LJ}(\lambda_1) =  U_\mathrm{P} + \lambda_1\left( U_\mathrm{P+X_{LJ}} - U_\mathrm{P} \right) \, ,
\end{equation}
where $\lambda_1$ is the scaling parameter, $U_\mathrm{P}$ is the potential of the pure polymer, and $U_\mathrm{P+X_{LJ}}$ is the potential of the pure polymer plus ion-polymer Lennard-Jones interactions.
Standard $\lambda$-dependent soft-core Lennard-Jones potentials, as implemented in LAMMPS with $n=1$ and $\alpha_{\mathrm{LJ}}=\mathrm{0.5}$, were used for all ion-polymer interactions to smooth the potential energy function.\cite{Beutler:1994aa} When $\lambda_1=0$, the ion and polymer are non-interacting, and when $\lambda_1=1$, the ion-polymer Lennard-Jones terms fully contribute to the potential energy. A second scaled potential was used to subsequently introduce the polymer-ion electrostatic interactions
\begin{equation}
\label{electrostatics_phase}
%  U_\mathrm{\lambda_2,C}(\lambda_2) =  U_\mathrm{P+X_{LJ}} + \lambda_2\left( U_\mathrm{P+X_{LJ}+X_{C}} - U_\mathrm{P+X_{LJ}} \right)
  U_\mathrm{C}(\lambda_2) =  U_\mathrm{P+X_{LJ}} + \lambda_2\left( U_\mathrm{P+X_{LJ}+X_{C}} - U_\mathrm{P+X_{LJ}} \right) \, ,
\end{equation}
where $U_\mathrm{P+X_{LJ}+X_{C}}$ is the potential of the polymer plus all polymer-ion interactions. The potential in Eq. \eqref{electrostatics_phase} was implemented by using $\lambda$-scaled charges on the ion. When $\lambda_2=0$, the ion and polymer interact only through the Lennard-Jones potential, and when $\lambda_2=1$, both the ion-polymer electrostatic and Lennard-Jones terms fully contribute to the potential energy.
The total solvation free-energy was obtained by
\begin{equation}
\label{total_delta_g}
%\Delta G=\int_{0}^{1}\left\langle \frac{\mathrm{d}U_\mathrm{LJ}}{\mathrm{d}\lambda_1} \right\rangle_{\lambda_2=0} \mathrm{d}\lambda_1 + \int_{0}^{1}\left\langle \frac{\mathrm{d}U_\mathrm{C}}{\mathrm{d}\lambda_2} \right\rangle_{\lambda_1=1} \mathrm{d}\lambda_2\,.
\Delta G_{\mathrm{TI}} = \int_{0}^{1}\left\langle \frac{\mathrm{d}U_\mathrm{LJ}}{\mathrm{d}\lambda_1} \right\rangle \mathrm{d}\lambda_1 + \int_{0}^{1}\left\langle \frac{\mathrm{d}U_\mathrm{C}}{\mathrm{d}\lambda_2} \right\rangle \mathrm{d}\lambda_2\,.
\end{equation}
The brackets in Eq. \eqref{total_delta_g} indicate an ensemble average, and the approximation has been made that the $P\Delta V$ contribution to the free energy change can be safely neglected. 
The integrals in Eq. \eqref{total_delta_g} were evaluated numerically using the trapezoidal rule, with $\lambda_1$ and $\lambda_2$ incremented in steps of 0.1 (twenty-one steps total, eleven for the Lennard-Jones phase and eleven for the electrostatics, less one redundant step connecting the two phases). 
The system was allowed to equilibrate for 100 ps at each $\lambda$-step, then an additional 100 ps of dynamics were used for calculating the necessary derivatives. The derivatives in Eq. \eqref{total_delta_g} were calculated by finite-difference. At endpoints, forward or backward finite-difference was used, at all other points the central difference was used with a $\lambda$-step of 0.01 to evaluate the derivative. In the case of TFSI${}^-$, an additional free energy contribution associated with removing the intramolecular electrostatics must be computed. Free-energy perturbation was used to evaluate this contribution from a ten ns MD trajectory of a single TFSI${}^-$ molecule in vacuum. The reported $ \Delta G_\mathrm{TI} $ values were calculated as the average over all ion-insertion trajectories, with errors in the mean estimated by bootstrap resampling (5 million samples).\cite{Chapman:1993} 

It is common for borane centers to undergo changes in bonding hybridization and associated  structural rearrangements upon anion complexation.\cite{Onak:1975} Since these effects are not captured by the force-field used in this work, it is anticipated that the MD thermodynamic integration calculations significantly underestimate the anion solvation free energies in the Lewis-acidic polymers. Using \emph{ab initio} geometry optimizations and energy calculations, we thus include a free-energy-perturbation correction to the solvation free energy (for both the Lewis-acidic polymers and PEO) associated with local structural relaxation during anion complexation,
\begin{equation}
\label{delta_g_corr}
\Delta G_\mathrm{corr} = \Delta G_\mathrm{rel,anion} - \Delta G_{\mathrm{rel,neat}},
% \beta^{-1} \mathrm{ln}(\left\langle e^{-\beta \Delta U_\mathrm{Neat}} \right\rangle )  - \beta^{-1} \mathrm{ln}(\left\langle e^{-\beta \Delta U_{\mathrm{Anion}}} \right\rangle ),
\end{equation}
where
\begin{equation}
\Delta G_\mathrm{rel,anion} = -\beta^{-1} \mathrm{ln}\left\langle e^{-\beta \Delta U_{\mathrm{rel,anion}}} \right\rangle
\label{delta_g_corr1}
\end{equation}
accounts for the free-energy change in the anion-polymer system upon relaxation,
and 
\begin{equation}
\Delta G_{\mathrm{rel,neat}} = -\beta^{-1} \mathrm{ln}\left\langle e^{-\beta \Delta U_\mathrm{rel,neat}} \right\rangle
\label{delta_g_corr2}
\end{equation}
accounts for the corresponding effect in the neat polymer.
The angle brackets in Eqs. \eqref{delta_g_corr1} and \eqref{delta_g_corr2} correspond to ensemble averaging over the MD configurations, and the configuration-specific relaxation energies $\Delta U_{\mathrm{rel,anion}}$ and $\Delta U_{\mathrm{rel,neat}}$ are obtained from quantum chemistry calculations at the PBE-D3/def2-SVP level of theory, as described below.

The $\Delta U_{\mathrm{rel,anion}}$ terms in Eq. \eqref{delta_g_corr1} were calculated by performing constrained geometry optimizations of the anion solvation structures. For each anion in each polymer, ten solvation snapshots were extracted at 100-picosecond intervals from the dilute ion concentration MD trajectories. Each solvation snapshot consisted of the anion and all polymer atoms within $r_\mathrm{outer} = 5 \,\mathrm{\AA}$ of any anion atom. The value of $r_\mathrm{outer}$ was chosen to be large enough to include the first solvation shell for all combinations of polymer and anion ($r_\mathrm{inner} = 4.5 \, \mathrm{\AA}$, see Fig. S1) plus a buffer region of $0.5 \, \mathrm{\AA}$.
The united-atom moieties and the terminal atoms of polymer fragments in the solvation snapshot were then hydrogenated, and a constrained geometry optimization of the added hydrogens was performed. 
Subsequently, a constrained geometry optimization was performed in which all binding atoms (carbon for PEO and boron for all Lewis-acidic polymers) within $r_\mathrm{inner}$ of any anion atom and polymer atoms within two bonds of those binding atoms were relaxed, while the positions of all anion atoms and remaining polymer atoms were constrained. 
For instances where these cutoff values return a polymer fragment without any constrained atoms, a position constraint was placed on the atom in the fragment farthest from the anion. 
The relaxation energy, $\Delta U_\mathrm{rel,anion}$, is obtained from  the single-point energy difference for the system before and after the constrained geometry optimization. 
%change in the quantum chemical energy upon relaxation was calculated as the single-point energy difference between the initial and final geometries of the solvation geometry optimization.
For each combination of polymer and anion, the average number of polymer binding atoms that were within $r_\mathrm{inner}$, $N_\mathrm{poly}$, is determined, and this value is used in the protocol for the calculation of $\Delta U_{\mathrm{rel,neat}}$.
%$N_\mathrm{poly}$ corresponds to the average number of polymer binding atoms that were within $r_\mathrm{inner}$ for a given combination of polymer and anion and is used in the protocol for the calculation of $\Delta U_{\mathrm{rel,neat}}$.}

The $\Delta U_{\mathrm{rel,neat}}$ terms in Eq. \eqref{delta_g_corr2} were similarly calculated by performing constrained geometry optimizations of snapshots obtained from neat-polymer MD trajectories. 
The procedures for sampling and optimizing the neat polymer snapshots were identical to the solvation geometries, except that the neat polymer snapshots included all polymer atoms within $r_\mathrm{outer} = 7.5 \, \mathrm{\AA}$ of the center of the simulation box; 
this larger value of $r_\mathrm{outer}$ was used in the neat-polymer calculations to account for the reduced density of binding atoms in the neat-polymer snapshots relative to the anion-containing snapshots.  
During the geometry relaxation, only the $N_\mathrm{poly}$  binding atoms closest to the center of the simulation box and the polymer atoms within two bonds of those binding atoms were allowed to relax, where $N_\mathrm{poly}$ is defined above; 
this ensures that the same average number of binding atoms are relaxing in calculating both $\Delta U_\mathrm{rel,anion}$ and  $\Delta U_{\mathrm{rel,neat}}$.
The neat relaxation energy, $\Delta U_\mathrm{rel,neat}$, is obtained from the single-point energy difference for the system before and after the constrained geometry optimization.

The final expression used to evaluate the solvation free energy for an anion was
\begin{equation}
\label{total_delta_g_anion}
\Delta G_\mathrm{S} = \Delta G_\mathrm{TI} + \Delta G_\mathrm{corr}.
\end{equation}
Table S1 includes the values of $\Delta G_\mathrm{corr}$ for all anions in all polymers. For \ce{Li+} in all polymers, $\Delta G_\mathrm{S} = \Delta G_\mathrm{TI}$. 

\subsection*{Salt Lattice Energies}
 The salt lattice energies for LiCl and LiTriflate presented in Fig. 5b were taken from reference 16. No experimental reference exists for LiTFSI, so the lattice energy was estimated using a modified Kapustinki approach\cite{Jenkins:1998aa} developed by Jenkins and a volume of 136.1 ${\text{\AA}}^3$ for TFSI calculated based on the DFT optimized structure at the B3LYP-D3/def2-TZVP level.
 
\bibliographystyle{naturemag}

\newpage
\begin{table}[htp!]
\centering
\small
\caption{Apparent ion diffusivities and monomer-unit diffusivities\textsuperscript{a} calculated for each of the polymer electrolytes in the dilute-ion regime.\textsuperscript{b}}
\begin{tabular}{c|ccccc}
\hline \hline
Polymer & $D_\mathrm{Li}^{\mathrm{app}}$ & $D_\mathrm{Cl}^{\mathrm{app}}$ & $D_\mathrm{triflate}^{\mathrm{app}}$ & $D_\mathrm{TFSI}^{\mathrm{app}}$ & $D_\mathrm{seg}^{\mathrm{app}}$ \\
\hline
PEO        &    $0.79(6)$      &   $22(2)$     &  $11(1)$       &   $5.9(6)$  &  $1.05(2)$    \\
CBC        &    $2.7(3)$      &   $2.4(3)$   &  $7.1(7)$        &   $17(1)$    &  $1.63(2)$    \\
CBCC      &    $0.16(1)$    &  $0.66(6)$  &  $1(2)$          &  $0.8(1)$    &  $0.79(1)$  \\
HBC        &    $ 0.56(9)$   &  $1.3(3)$   &  $1.3(2)$         &   $1.3(1)$  &  $2.07(6)$   \\
HBCC      &    $6(1)$         &  $4.0(7)$   &  $16(3)$         &    $16(3)$   &  $1.37(2)$   \\
\hline \hline
%\multicolumn{6}{l}{\textsuperscript{b}\footnotesize{Units of $10^{-7} \, \mathrm{cm^2 \cdot s^{-1}}$}}\\
%\multicolumn{6}{l}{\textsuperscript{e}\footnotesize{Other problems: sig digs, statement about parentheses, a and b subscripts not clearly applied. Units only stated in b?}}\\
%\multicolumn{6}{l}{\textsuperscript{f}\footnotesize{Numbers in parentheses indicate the statistical uncertainty in the last reported digit.}}\\
\end{tabular}

\begin{flushleft}
\textsuperscript{a}\footnotesize{Calculated based on the average MSD of oxygen or boron in each polymer}\\
\textsuperscript{b}\footnotesize{For all cases, apparent diffusivity is based on the MSD at $t = 150$~ns, as described in the methods section.  Units of $10^{-7} \, \mathrm{cm^2 \cdot s^{-1}}$ are reported. Numbers in parentheses indicate the statistical uncertainty in the last reported digit.}
\end{flushleft}
\end{table}

%%%%%%%%%%%%%%%%%%%%%%%%%%%%%%%%%%%%%%%%%%%%%%%%%%%%%%%%%%%%%%%%%%%%%
%% Figures
%%%%%%%%%%%%%%%%%%%%%%%%%%%%%%%%%%%%%%%%%%%%%%%%%%%%%%%%%%%%%%%%%%%%%
\begin{figure}
  \centering
    \includegraphics[width=170mm]{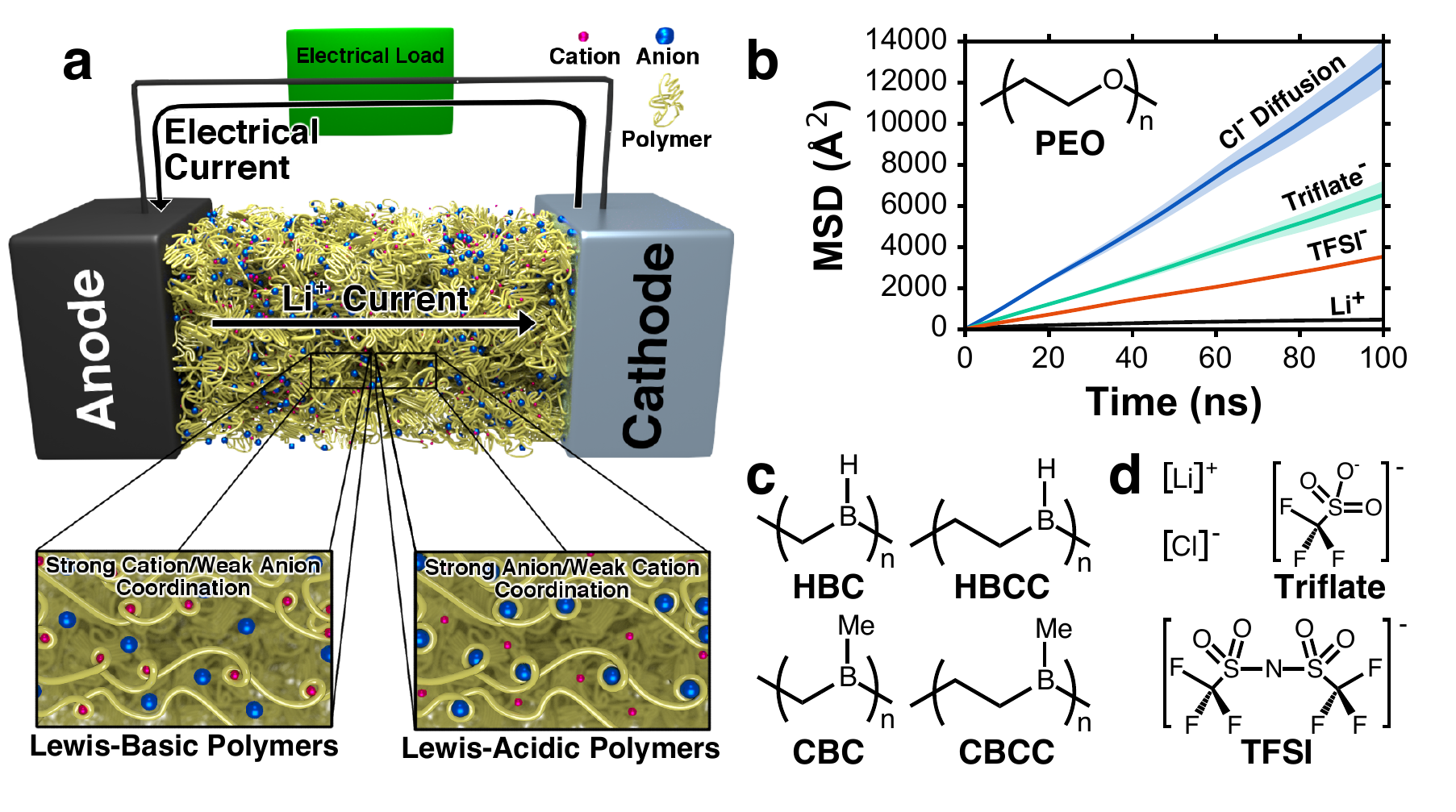}
  \caption{ Overview of the new polymer electrolyte concept, motivating problem, and materials. (a) Schematic depiction of solid polymer electrolyte operation in Lewis-basic polymers and Lewis-acidic polymers. Cations are preferentially complexed in the conventional Lewis-basic environment, whereas anion complexation is promoted in Lewis-acidic environments. (b) Mean-squared displacement of \ce{Li+} and several anions in PEO in the dilute-ion regime, based on MD trajectories at 400 K. The asymmetric transport of anions and cations in PEO is reflected in the higher diffusivities of all anions relative to \ce{Li+}. Average values are plotted as solid lines and standard errors are plotted as shaded regions. (c) Chemical structures of the Lewis-acidic polymers simulated in this study. (d) Chemical structures of the ions simulated in this study.}
  \label{fgr:Figure1}
\end{figure}

\begin{figure}
  \centering
  \includegraphics[width=170mm]{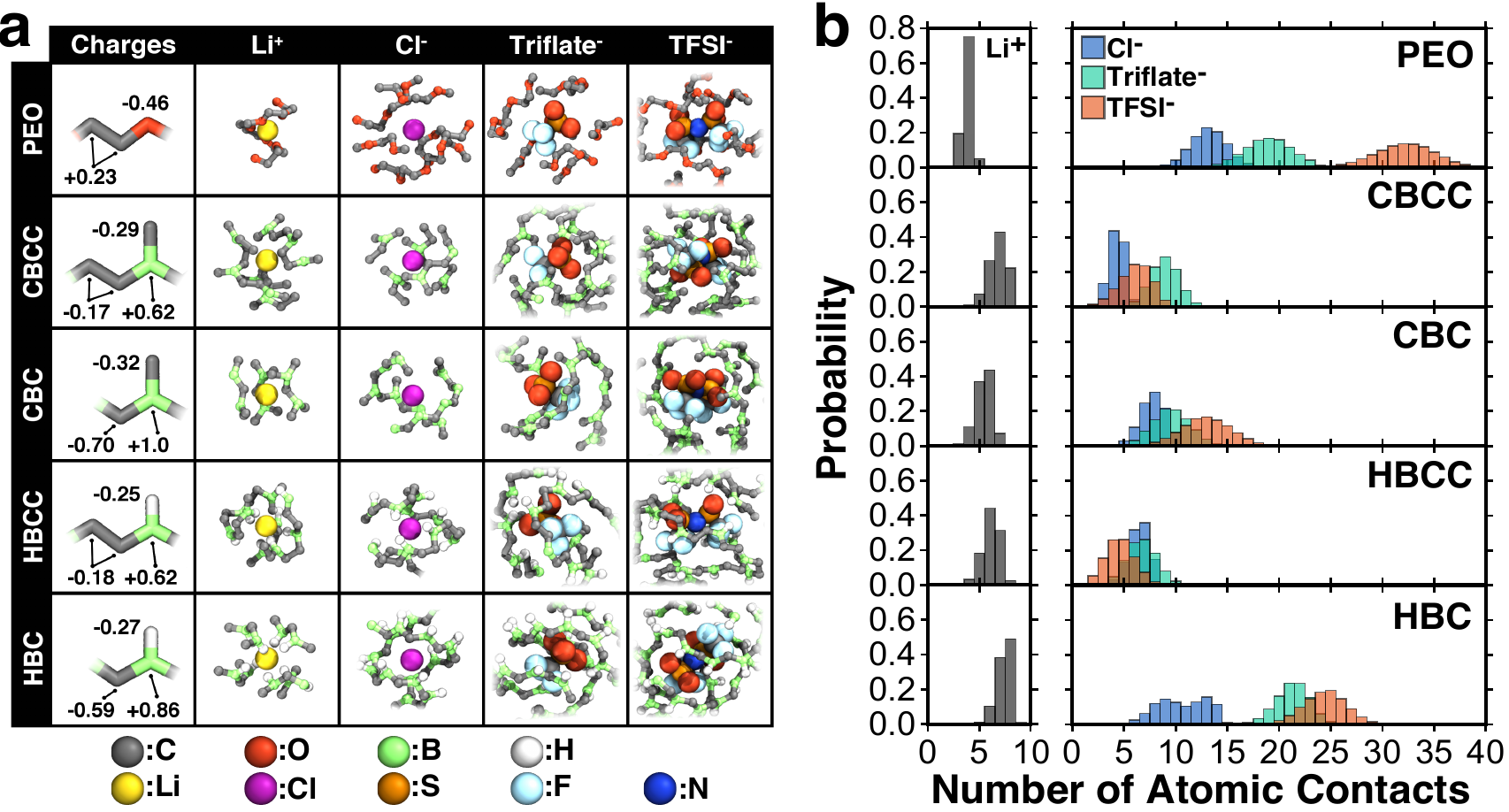}
  \caption{Contrasting ion coordination behavior in Lewis-basic and Lewis-acidic polymers. (a) Representative coordination structures of each ion in each polymer. The CHELPG derived partial charges used in the MD simulations are shown in the left column. (b) Histograms of the number of polymer-ion contacts averaged across the simulations. All data is derived from MD trajectories at 400 K in the dilute-ion regime.}
  \label{fgr:Figure2}
\end{figure}

\begin{figure}
  \centering
  \includegraphics[width=88mm]{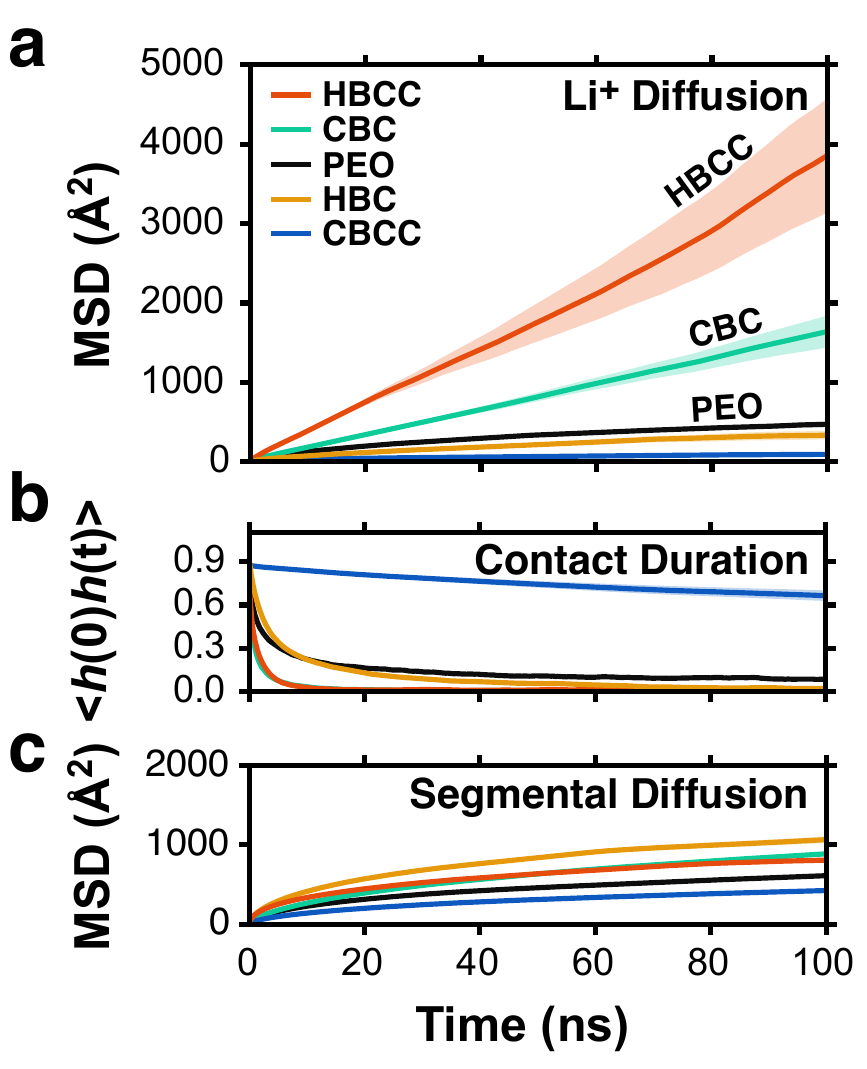}
  \caption{Contrasting \ce{Li+} transport behavior in Lewis-basic and Lewis-acidic polymers. (a) Comparison of the mean-squared displacement (MSD) of \ce{Li+} in all polymers. (b) Contact durations for \ce{Li+} and the predominant binding atoms in each polymer (see Figure 2a, O for PEO;   CH${}_3$ for CBC; CH${}_2$ for CBCC; and H for HBC and HBCC). (c) MSD of the polymer monomer units.  All data is derived from MD trajectories at 400 K in the dilute-ion regime. Average values are plotted as solid lines and standard errors are plotted as shaded regions.}
  \label{fgr:Figure3}
\end{figure}

\begin{figure}
  \centering
  \includegraphics[width=88mm]{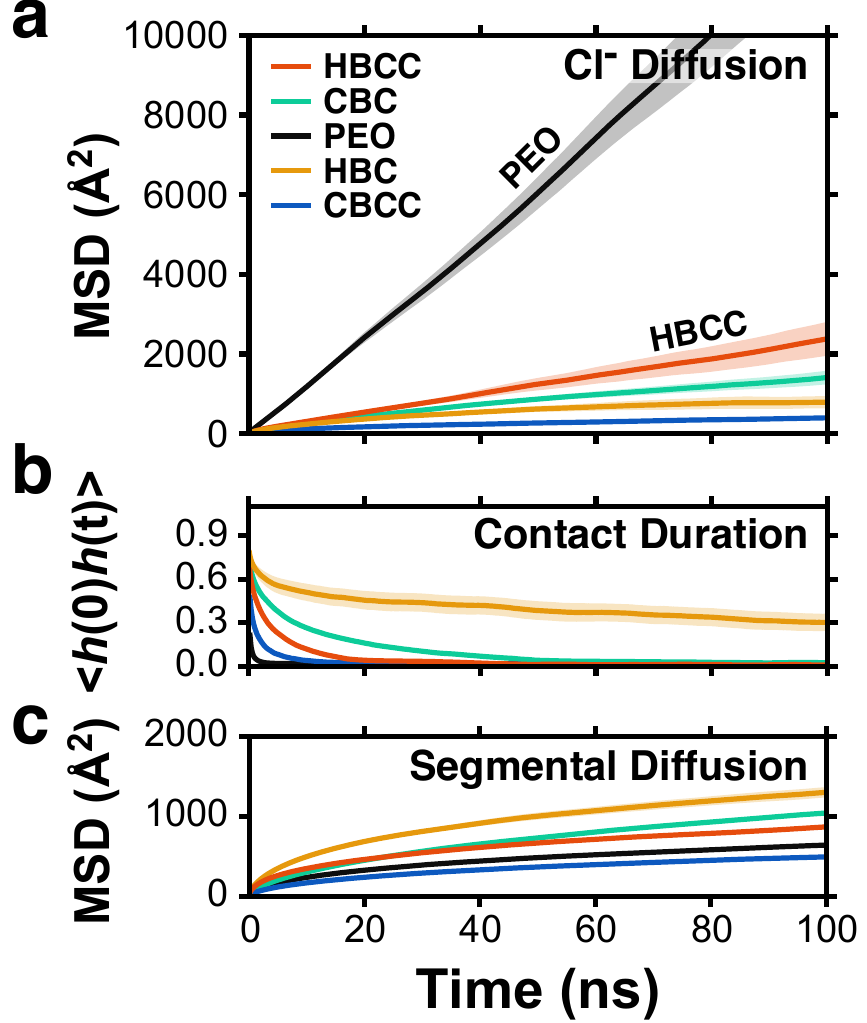}
  \caption{Contrasting \ce{Cl-} transport behavior in Lewis-basic and Lewis-acidic polymers. (a) Comparison of the mean-squared displacement (MSD) of \ce{Cl-} in all polymers. (b) Contact durations for \ce{Cl-} and its predominant binding atom in each polymer (see Figure 2a, C for PEO and B for all Lewis-acidic polymers). (c) MSD of the polymer monomer units.  All data is derived from MD trajectories at 400 K in the dilute-ion regime. Average values are plotted as solid lines and standard errors are plotted as shaded regions.}
  \label{fgr:Figure4}
\end{figure}

\begin{figure}
  \centering
  \includegraphics[width=88mm]{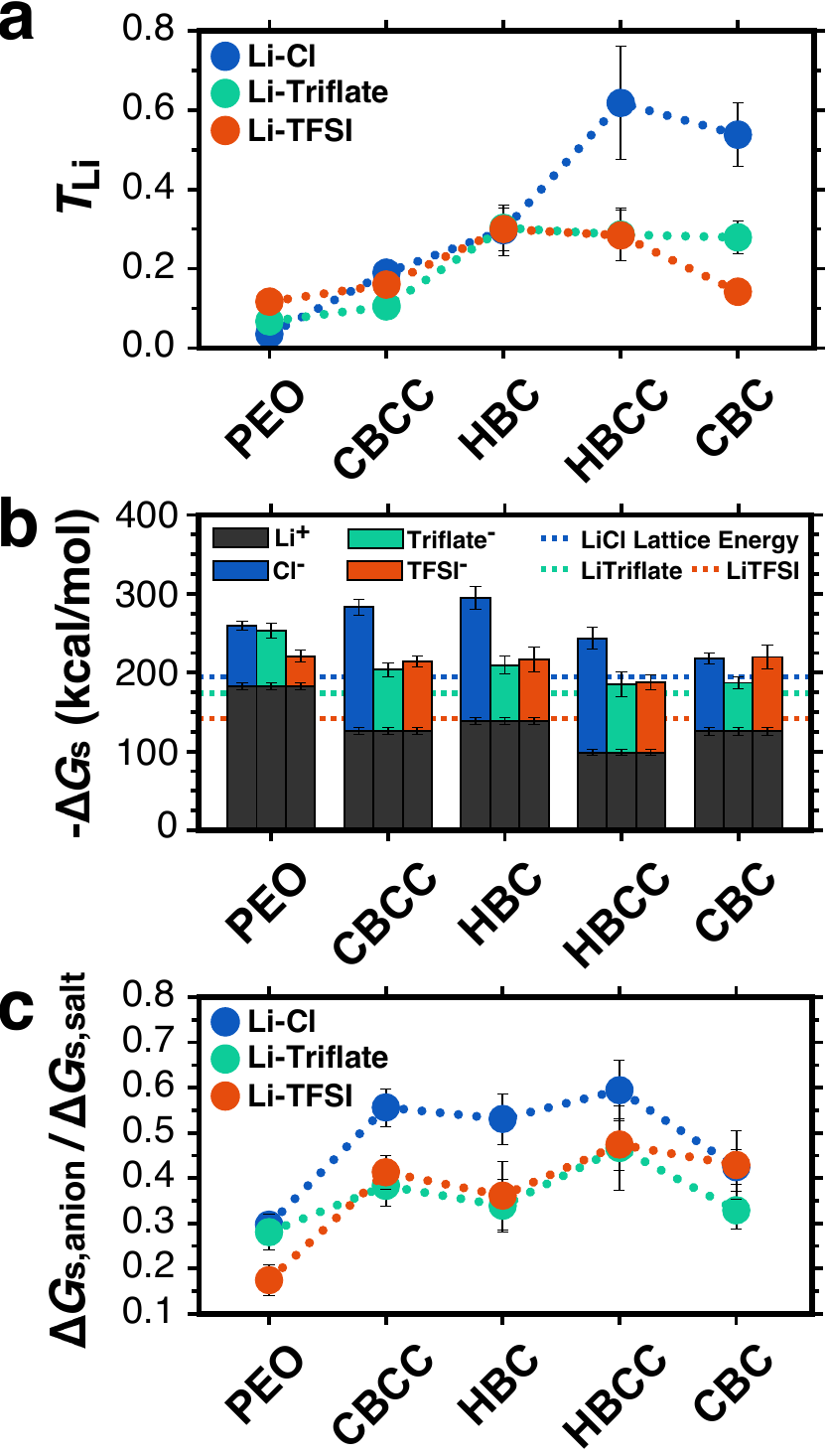}
  \caption{Comparison of $T_{\mathrm{Li}}$ and ion solvation free energies in Lewis-basic and Lewis-acidic polymers. (a) Transference numbers for all salts in the dilute-ion regime. Dotted lines are a guide to the eye. (b) Solvation free energies for each ion in each polymer and the lattice energies of the corresponding Li-salts (dotted lines). (c) Fractional contribution of each anion to the total salt solvation free energy for each polymer. All free energies are calculated via thermodynamic integration at 400 K in the dilute-ion regime. The anion solvation free energies have an additional contribution from the quantum chemical relaxation of the solvation structures calculated via free-energy perturbation.}
  \label{fgr:Figure5}
\end{figure}

\end{document}